\documentclass[10pt]{article}
\usepackage[letterpaper]{geometry}
\usepackage{hicss}
\usepackage{times}
\usepackage[none]{hyphenat}
\usepackage{url}
\usepackage{latexsym}
\usepackage{indentfirst}
\usepackage{graphicx}
\usepackage{array}
\usepackage{amsmath} 
\usepackage{pgfplots}
\usepackage{pgfplotstable}
\pgfplotsset{compat=1.13}
\definecolor{constructCluster}{HTML}{2B83BA}
\definecolor{dbscan}{HTML}{ABDDA4}
\usepackage{footnote}
\makesavenoteenv{tabular}
 \usepackage{comment} 
\usepackage{natbib}
\title{Recognizing Similar Crises through the Application of Ontology-based Knowledge Mining}

\author{Ngoc Luyen Le$^{1,2}$\\
	{ngoc-luyen.le@hds.utc.fr} \\\And
	Marie-Hélène Abel$^1$\\
	{marie-helene.abel@hds.utc.fr}\\\And
	Elsa Negre$^3$\\
	{elsa.negre@dauphine.fr}
	\AND 
	\hspace{-1.1cm}$1$: Université de technologie de Compiègne, CNRS, Heudiasyc, CS 60319 - 60203 Compiègne Cedex, France.
	\\
	\hspace{-7.55cm}$2$: Vivocaz, 8 B Rue de la Gare, 02200, Mercin-et-Vaux, France.
	\\
	\hspace{-1.8cm}$3$: Paris-Dauphine University, PSL Research Universities, CNRS UMR 7243, LAMSADE, Paris, France.\\
}
\date{}

\begin{document}
	\maketitle
	\begin{abstract}
		Recognizing and learning from similar crisis situations is crucial for the development of effective response strategies. This study addresses the challenge of identifying similarities within a wide range of crisis-related information. To overcome this challenge, we employed an ontology-based crisis situation knowledge base enriched with crisis-related information. Additionally, we implemented a semantic similarity measure to assess the degree of similarity between crisis situations. Our investigation specifically focuses on recognizing similar crises through the application of ontology-based knowledge mining. Through our experiments, we demonstrate the accuracy and efficiency of our approach to recognizing similar crises. These findings highlight the potential of ontology-based knowledge mining for enhancing crisis recognition processes and improving overall crisis management strategies.
	\end{abstract}
	
	\subsubsection*{Keywords: Crisis Management, Knowledge Representation, Ontology, Semantic Similarity}
	
	\vspace{-0.2cm}
	\section{Introduction}\label{introduction}
	\vspace{-0.2cm}
	In the field of crisis management, the ability to recognize and learn similar crisis situations is of paramount importance for devising effective response strategies. By identifying similarities between different crises, decision-makers can leverage existing knowledge and experiences to inform their decision-making processes, enhance response capabilities, and mitigate the potential impact of crises \citep{bundy2017crises, t2023teaching}. Furthermore, the recognition of similar crises enables the identification of relevant actions and effective communication strategies specific to each crisis situation \citep{negre2013towards}. Understanding the similarities between crises is crucial for decision-makers as it allows them to gain insights into the limitations of traditional crisis management practices and identify the most appropriate interventions and measures based on the experiences and outcomes of past crisis situations \citep{boin2007preparing}. This enables more targeted and tailored responses, optimizing the allocation of resources and minimizing potential disruptions.  However, the task of identifying such similarities across a wide range of crisis-related information and scenarios can often be daunting and time-consuming \citep{boin2020learning}.

	 
	 To address this challenge, ontology-based knowledge mining techniques offer a promising solution for structuring and organizing crisis-related information and scenarios. {Ontology, as a structured framework for modeling knowledge within specific domains, formalizes concepts and relationships, thereby enhancing data understanding, retrieval, and integration \citep{gruber1993translation}.} By capturing the relationships, attributes, and properties of various crisis elements, an ontology-based knowledge base becomes a valuable resource for identifying patterns and similarities within the domain of crises. The use of ontology-based knowledge mining in crisis management offers several advantages. It allows for the integration and consolidation of diverse data sources, including historical records, expert knowledge, and data feeds, to build a comprehensive knowledge base. 
	 Through modeling, reasoning, and inference techniques, ontology-based knowledge mining can uncover meaningful connections and associations, enabling the identification of similar crisis situations and providing decision-makers with valuable insights into crisis situations.
	 \citep{abiteboul_2011}. 
	
	The recognition of similar crises necessitates the development of a systematic approach, which forms the core of our investigation in this study. In this paper, we focus on recognizing similar crises through the application of ontology-based knowledge mining. Our methodology involves the utilization of an ontology-based crisis situation knowledge base, enriched with relevant crisis-related information. Additionally, we employ a semantic similarity measure to assess the degree of similarity between different crisis situations. By combining these components, we aim to enhance the accuracy and efficiency of crisis recognition processes, thereby enabling more effective crisis management and response strategies.
	
	The remainder of this paper is organized as follows: Section \ref{related_work} introduces works from the literature on which our approach is based. Section \ref{approach} presents our main contributions, outlining the methodology for recognizing similar crises through ontology-based knowledge mining and the use of semantic similarity measure. In section \ref{experiments}, we provide experiments about identifying and calculating the crises similarities based on our developed ontology. Finally, we conclude and discuss the perspectives.
	
	
	\vspace{-0.2cm}\section{Related Work}\label{related_work}\vspace{-0.2cm}
	In this section, our focus is on exploring the utilization of ontologies for knowledge representation and investigating different approaches for modeling crisis management and identifying similar crises. We delve into the use of ontologies as a foundation for structuring and organizing crisis-related knowledge. \vspace{-0.2cm}
	\subsection{Crisis-Related Information Management}\vspace{-0.2cm}
	In this paper, we focus on crisis-related information management, specifically addressing the various aspects and challenges related to the effective structuring, organization, and utilization of information during three critical phases: pre-crisis, crisis response, and post-crisis. The utilization of ontology-based knowledge representation serves as our primary approach to advance the organization and reusability of crisis-related information. 
	
	Crisis-related Information Management focuses on the efficient collection, organization, and utilization of information during crisis situations. It encompasses a range of activities and processes aimed at acquiring, processing, analyzing, and disseminating information to support decision-making and response efforts \citep{meesters2021crisis,Bowen2020}. Various studies have explored the application of ontologies in different aspects and situations of crisis-related information management. For instance, in the study by \cite{de2019creative}, an emergency management ontology was created with a focus on knowledge related to smart cities and crises arising from natural or anthropogenic events. 
	 Furthermore, \cite{elmhadhbi2020promes} introduced the \textit{POLARISCO} ontology, which captures knowledge from various stakeholders at both the commandment and operational levels of disaster response.
	\citet{benaben2016metamodel,benaben2020ai} introduced the \textit{COSIMMA} metamodel which offers a systematic abstract representation of knowledge in the field of collaborative situations in crisis management. The metamodel encompasses four interconnected dimensions: context, partners, objectives, and behavior. In the study by \cite{le:hal-03832226}, the authors proposed the development of a lightweight crisis management domain ontology by leveraging the inheritance of the ISyCri meta-model \citep{benaben2008metamodel}. The ontology incorporates concepts related to crisis description, affected individuals, and available resources. In summary, the existing ontologies in the field of crisis management are highly reliant on specific crisis scenarios, each containing unique knowledge and information pertaining to the particular type of crisis and/or geographical area. However, these ontologies lack a systematic analysis of the model and organization for the effective management of general crisis-related information.
	  
	The structure and organization of crisis-related information enable the storage of specific characteristics related to each individual crisis. Developing a comprehensive ontology for crisis-related information management facilitates the storage, retrieval, and reuse of information, allowing for valuable insights and experiences to be extracted regarding actions, utilized resources, and communication methods. In the upcoming section, we will delve into existing research on leveraging ontology for the recognition of similar crises.
	 \vspace{-0.2cm}\subsection{Similar Crisis Identification}\vspace{-0.2cm}
	 
	In order to recognize similar crises, the application of the Case-Based Reasoning (CBR) approach can be instrumental in finding similar cases. Indeed, CBR refers to a problem-solving methodology that leverages past experiences or cases to solve new problems or make decisions. In the context of crisis management, CBR involves comparing the characteristics and attributes of different crisis situations and identifying similarities based on their shared features. By analyzing and comparing relevant factors such as the nature of the crisis, the affected population, response strategies, and outcomes, a CBR system can identify similar crisis scenarios. 
	 In recent years, several studies have employed the CBR approach to tackle the problems of identifying similar crises and planning crisis responses. For instance, in \cite{yu2018risk}, the authors specifically examined risk response in the field of critical infrastructure protection, with a particular emphasis on natural disasters. The research underscores the significance of historical cases as valuable resources that enable timely decision-making in the development of response strategy plans. 
	
	The identification of similar crises can be effectively achieved by utilizing knowledge-based recommender systems, which have demonstrated their value in numerous domains, including crisis management. By analyzing historical crisis data and patterns, knowledge-based recommender systems can identify similarities and patterns among different crises, enabling the identification of similar crisis scenarios. 
	For example, in the study conducted by \cite{negre2013towards}, the author proposed a methodology for identifying similar actions and leveraging knowledge gained from past experiences to support decision-makers in crisis management. The objective is to enhance crisis management by providing decision-makers with valuable insights and lessons learned from similar crisis situations. In \cite{le:hal-03832226}, a constraint-based recommender system was proposed to optimize the deployment of citizen vehicle resources in the context of crisis management. The objective was to develop a system that not only recommends the most suitable deployment strategies but also enhances the overall efficiency of utilizing citizen vehicles during crisis situations.
	
	CBR demonstrates a higher level of adaptability when it comes to novel crises, as it leverages case reuse and adaptation. In contrast, knowledge-based recommender systems may face challenges when dealing with new or rare crisis scenarios due to limited historical data. Additionally, CBR is typically domain-specific, specifically tailored to crisis management, while knowledge-based recommender systems have broader applicability across various domains. However, both CBR and knowledge-based recommender systems share similarities in their approach to identifying similar crisis situations. They both rely on existing knowledge and patterns, utilizing past cases or experiences to provide recommendations or solutions. Moreover, both approaches employ similarity measures that take into account crisis characteristics such as type, context, and impact to determine similarity.  In the following section, we delve into the exploration of our approach for identifying similar crisis situations.
	 
	\vspace{-0.2cm}\section{Our approach}\label{approach}\vspace{-0.2cm}
	In this section, we present our approach, which involves processing cycles for recognizing similar crisis situations, constructing an ontology-based crisis situation knowledge base, and developing a method for semantic similarity measurement using the ontology-based crisis situation knowledge base.
	\vspace{-0.2cm}\subsection{Process Cycle Formulation}\vspace{-0.2cm}
	
\begin{figure}[thb]
		\centering
		\includegraphics[width=0.8\linewidth]{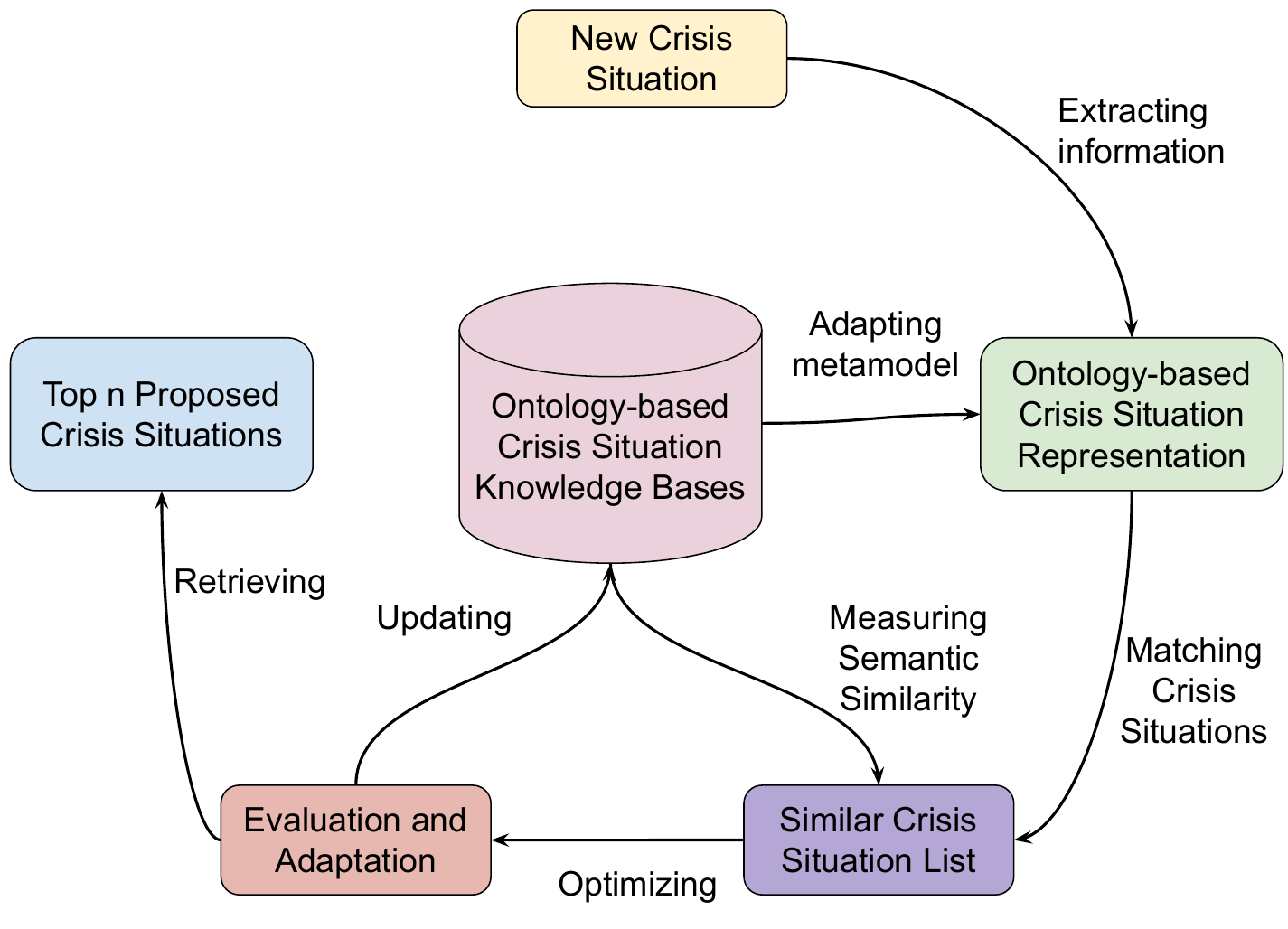}
		\caption[justification=justified]{A process cycle for recognizing similar crisis situations.}
		\label{fig_01}       
		\vspace{-0.3cm}
	\end{figure}

	We present our process cycle for recognizing similar crisis situations, as illustrated in Figure \ref{fig_01}. At the onset of a new crisis situation, the relevant information pertaining to the event is meticulously extracted and transformed into an ontology-based representation, ensuring a standardized and structured format for further analysis. The construction of the new crisis situation adopts a metamodel derived from the ontology-based crisis situation knowledge base, which serves as a foundation for accurately reconstructing the key elements and attributes of the crisis.
	Moving forward, the process entails the critical step of matching and measuring the semantic similarity between the newly identified crisis situation and the existing crisis knowledge base built upon the ontology. This comparison is aimed at determining the degree of resemblance and relevance between the new crisis situation and previously encountered cases. By employing semantic similarity measurement techniques, a comprehensive list of the most pertinent and similar crisis situations is generated, offering decision-makers valuable insights and reference points for their decision-making process.
	Subsequently, the full information of the identified similar crisis situation cases is thoroughly evaluated and adapted to address the specific needs and requirements of decision-makers in their endeavor to mitigate the crisis. This evaluation involves a careful examination of the lessons learned, successful strategies, and best practices employed in the similar crisis situations, with the aim of providing practical guidance and solutions for problem-solving in the current crisis scenario.
	To ensure the currency and effectiveness of the crisis response system, the information obtained from the similar crisis situations can be integrated and updated in the ontology-based crisis situation knowledge base, enriching the repository with valuable experiences and insights. This ongoing refinement process ensures that the knowledge base remains up-to-date and capable of effectively supporting future crisis management efforts.
	In the final stage of the process cycle, a curated selection of the top n crisis situations, based on their relevance and similarity to the current crisis situation, is recommended to decision-makers. This curated list serves as a valuable resource, aiding decision-makers in their strategic planning, resource allocation, and overall crisis response efforts. By leveraging the knowledge and experiences encapsulated in the similar crisis situations, decision-makers can make well-informed and effective decisions to address the challenges posed by the crisis at hand.
	
	Building upon the general process cycle, this study primarily focuses on two key tasks that play a crucial role in recognizing similar crisis situations. The first task pertains to the construction of an ontology-based crisis situation knowledge base, which serves as the foundation for organizing and structuring crisis-related information. By developing this knowledge base, we aim to create a comprehensive repository of crisis situations, their characteristics, and associated knowledge, enabling efficient retrieval and analysis of relevant cases.
	The second task involves the implementation of a method for measuring semantic similarity between different crisis situations. This method serves as a crucial component in the process of identifying similar crisis situations from the knowledge base. By employing semantic similarity measures, we aim to capture the shared features, attributes, and context between crisis situations, enabling effective comparison and identification of similarities. In the next section, we will present the construction of the ontology-based crisis situation knowledge base.

	\vspace{-0.2cm}\subsection{Ontology-based Crisis Situation Knowledge Base}\vspace{-0.2cm}
	In this section, we delve deeper into the construction of the crisis-related information ontology, which forms an integral part of our ontology-based crisis situation knowledge base. This entails defining the ontology structure, identifying relevant crisis domains and sub-domains, and establishing the relationships and dependencies between different elements within the ontology. {In our work, using an ontology-based approach offers advantages such as semantic richness, automated reasoning, and flexible extensibility. Ontologies provide a structured and meaningful representation of knowledge, enhancing data analysis and decision-making through explicit relationships and inferences. They enable interoperability, contextual understanding, and advanced querying, accommodating domain-specific constraints. While simple database formats and RDF stores provide storage and querying capabilities, ontologies excel in capturing complex domain semantics, promoting shared understanding, and facilitating deeper insights from the data. The ontological choice allows to effectively manage the complexity of the domain, fulfill  the requirement for semantic reasoning, and achieve the goal of long-term scalability and interoperability.}
	
	\begin{figure*}[th]
	\centering
	\includegraphics[width=0.70\linewidth]{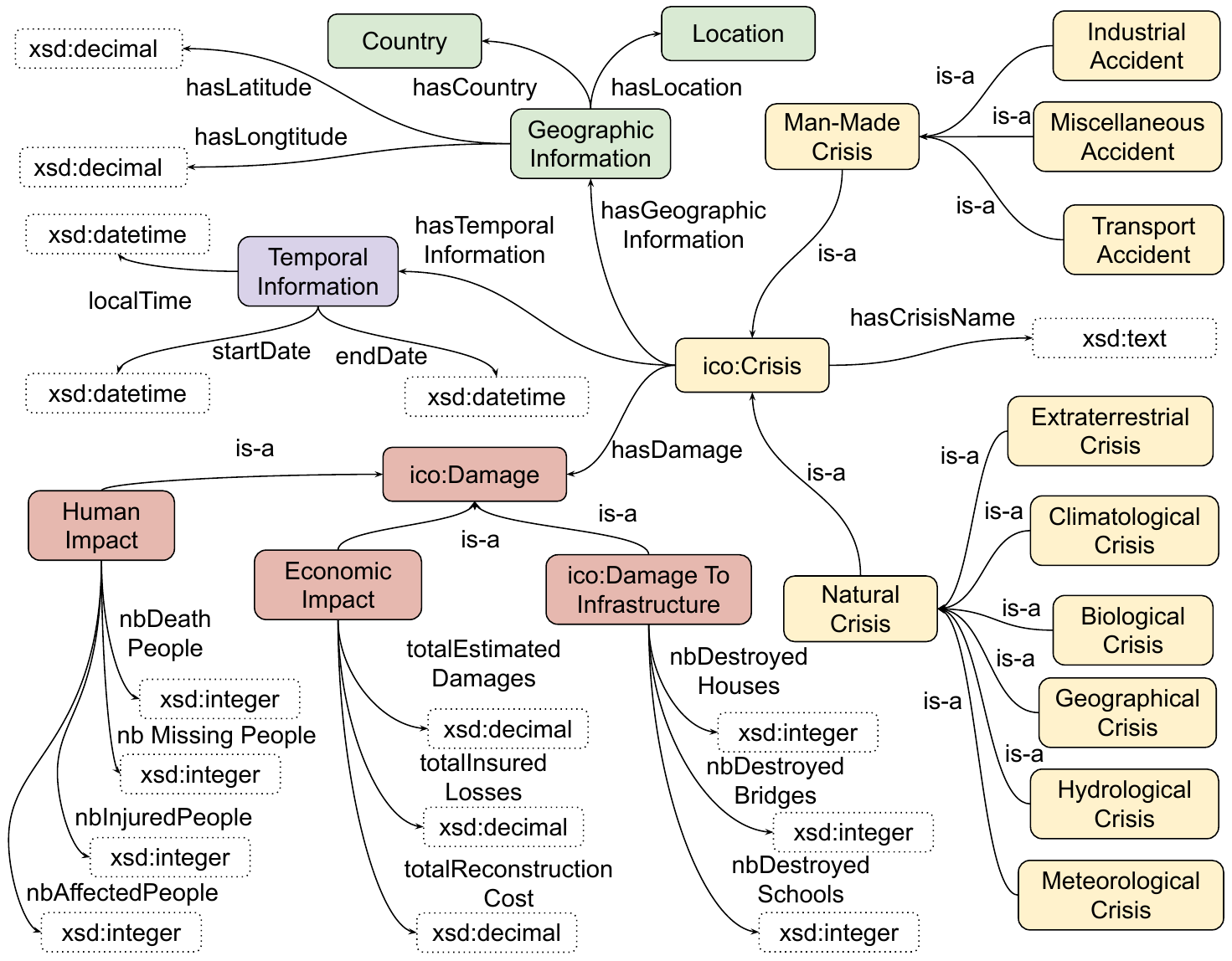}
	\vspace{-0.3cm}
	\caption{A dedicated ontology section for crisis-related information ontology included a hierarchy of crisis types (yellow boxes), damage (orange boxes), physical characteristics (blue boxes), geographical information (green boxes), and temporal entity (violet boxes), [``ico'': IsyCri meta-model, ``to'': Time Ontology].}
	\label{fig_02}       
	\vspace{-0.6cm}
\end{figure*}
	
	The process development of an ontology requires a systematic approach that involves gathering information, defining the scope, creating a conceptual model, defining the ontology schema, implementing the ontology, and evaluating and refining it as necessary. Therefore, we adopted the Agile Methodology for Ontology Development (AMOD) to develop our ontological model \citep{abdelghany2019agile}. In general, the AMOD methodology comprises three key phases: (i) the pre-game phase, where the goals, requirements, and scope of the ontology are clearly defined to provide a general overview; (ii) the development phase, where the project is divided into iterative sprints\footnote{A sprint is a short period of concentrated work within a project where a team works together to accomplish specific objectives or deliverables.}, each dedicated to essential tasks such as conceptualization, formalization, and integration, enabling a progressive advancement in the development of the ontology; and (iii) the post-game phase, which encompasses critical activities such as evaluating and maintaining the ontology to ensure its enduring reliability and effectiveness as time progresses.

	In line with the pre-game phase, our proposed ontology primarily aims to establish a standardized representation of crisis situation information and contextual details of crisis situations. This ontological model enhances information retrieval and adaptation within the crisis management system. The scope of the ontology encompasses a wide range of crisis situations and factors that contribute to the organization and structure of crucial information during crisis situations. It is designed to be flexible and adaptable, allowing for the incorporation of new crisis scenarios. Furthermore, our focus lies in creating a formal and structured representation of the crisis situation domain, with the objective of promoting a shared understanding and facilitating the use of standardized terminology among all stakeholders. By leveraging these valuable resources, our general aim is to improve the ability to locate relevant information and draw upon past experiences from similar crisis situations.

	The development of the crisis-related information ontology plays a critical role in improving the organization and management of complex crisis situations. In line with the development phase, this study focuses on constructing a comprehensive domain ontology that encompasses key concepts, relationships, and knowledge associated with crisis situations. The development of our domain ontology for crisis situations builds upon the foundations of the crisis management ontology proposed by \cite{le:hal-03832226} and the inheritance of the ISyCri meta-model \citep{benaben2008metamodel}. Indeed, the ISyCri meta-model bears many similarities to a meta-ontology as it encapsulates essential concepts for domain ontologies, establishing a close relationship between the two \citep{arru:hal-02276437}. Therefore, we have developed our ontology independently, without relying on a general top-level ontology like SUMO \citep{niles2001towards}, DOCLE \citep{gangemi2002sweetening}, or BFO \citep{arp2015building}. This deliberate choice allows us to create lightweight, customized, and context-specific ontologies that are specifically tailored for the crisis management domain. {The distinctiveness of the proposed ontology comes from how it's designed to match the needs of crisis-related data. Even though there might be similar classes in other ontologies as indicated in related work, in this ontology, their meanings and associations are customized to fit the specifics of crises. This ontology recognizes the varied and complex nature of crisis situations, aligning its classes and characteristics very closely with the details of crisis situations.}
	Exploring our developed ontology\footnote{Full version of crisis-related information ontology is published at \url{https://github.com/lengocluyen/OBCS-KB}.}, as depicted in Figure \ref{fig_02}, we extensively refine and enhance our proposed ontology by focusing on key entities: Crisis, Geographical Information, Temporal Entity, Physical characteristic, and Damage. 	
	 {In the paper, we did not explicitly differentiate between the terms `disaster' and `crisis' within the scope of ontological analysis. Indeed, these terms are often used interchangeably, and their distinction might not always be well-defined. Their usage can vary depending on factors such as the event's severity, the extent of impact, and the level of preparedness and response. }
	 Firstly, crises can be classified into various typologies based on factors such as the nature of risks, associated events, and public perceptions. For example, the information service of the French government\footnote{\url{www.gouvernement.fr/en}} categorizes major crises into terrorist, health, technological, natural, and cyber crises. Additionally, the Institute for Radioprotection and Nuclear Safety (IRSN)\footnote{\url{www.irsn.fr/EN}} classifies crises into individual, diffuse pollution, non-industrial collective, and technological crises related to specific sites and actions. In our work, we adopt the crisis classification proposed by the Centre for Research on the Epidemiology of Disasters (CRED) at the School of Public Health of the Université catholique de Louvain in Brussels, Belgium \citep{guha2016dat}. This classification organizes crises into a hierarchical structure based on crisis types, which further include natural and man-made crises.	 
	 Natural crises can be classified into several categories, including extraterrestrial, climatological, biological, geographical, hydrological, and meteorological crises. Each category is further divided into sub-categories. For example, meteorological crises can include extreme temperatures, fog, storms, and more. Geographical crises encompass earthquakes, mass movements, volcanic activity, and others. On the other hand, man-made crises are divided into industrial, miscellaneous, and transport accident crises. Similarly, these subcategories can be further categorized by other groups. For example, industrial accident crises may include explosions, gas leaks, collapses, oil spills, and more.
	Secondly, geographical information encompasses various aspects such as location, region, country, continent, as well as topography, specific longitude and latitude coordinates. Thirdly, temporal entity includes details regarding the starting and ending dates of the crisis, as well as the duration of the crisis. With temporal entity, the temporal entity, we utilize concepts and relationships from the time ontology\footnote{\url{https://www.w3.org/TR/owl-time/}}, ensuring consistency and reusability. Fourth, physical characteristic encompasses crucial information pertaining to the triggering origin of the crisis, as well as the magnitude scale and corresponding values associated with the event. Finally, the damages caused by the crisis event were categorized into human impact, economic impact, and damage to infrastructure. The human impact category focuses on statistics related to the number of affected, injured, missing, and deceased individuals. The economic impact category involves the total estimated damages, insured losses, and reconstruction costs. The damage to infrastructure category encompasses a wide range of damages, including destroyed or damaged houses, schools, commercial buildings, roads, bridges, and more.
	
	In accordance with the post-game phase of development, our attention is directed towards testing and evaluating the ontology that has been created. This involves conducting various types of tests and assessments to ensure the quality and effectiveness of the ontology. We can perform tests and evaluations on multiple aspects, including the overall structure of the ontology, the specific instances within it, and the retrieval of information from the ontology. By conducting these rigorous evaluations, we aim to validate the ontology's functionality, reliability, and its ability to meet the intended objectives of crisis-related information management.
	
	The implementation of the crisis-related information ontology involves the practical steps of creating and deploying the ontology within an application. This process entails converting the conceptual model of the ontology into a computer-readable format using ontology languages such as OWL (Web Ontology Language) and RDF (Resource Description Framework). In our study, we have implemented the ontology by leveraging OWL and utilizing the Protégé-OWL editor as a supportive tool \citep{musen2015protege}.
	In the upcoming section, we will investigate the approach employed to measure the semantic similarity of crisis situations, based on the utilization of the constructed ontology.
	\vspace{-0.2cm}
	\subsection{Semantic Similarity Measure}
	\vspace{-0.2cm}
	{We employ the term ``semantic similarity" rather than simply ``similarity" because it is based on the underlying meaning and relationships of the concepts represented within the ontology. 
		The term ``semantic" emphasizes the depth and context of the similarity calculation, reflecting the ontology's ability to capture the intricate semantic connections between crisis-related concepts.} The measure of semantic similarity involves the comparison of two sets of triples based on their individual elements, which can be categorized into quantitative and qualitative information. This process aims to assess the degree of similarity between the sets. By examining both the numerical and descriptive aspects of the triples, we can gain insights into their similarities and differences. Consequently, we utilize the semantic similarity measure by leveraging the ontology to identify similar crises in this section.
	
	Conventionally, an instance of crisis-related information, denoted as $C$, is commonly defined as a set of crisis, geographic information, temporal entity, physical characteristics and damage triples. This triple set is represented as follows:\vspace{-0.2cm}
	\begin{equation}\vspace{-0.2cm}
		C = \{a^C_1, a^C_2, ..., a^C_m\}
	\end{equation} where each $a^C_t$ represents a triple, specifically $a^C_t = \langle subject_t,$ $predicate_t,$ $object_t \rangle$. Alternatively, the triple $a^C_t$ in the context of Resource Description Framework is often used to represent a statement or assertion about a resource.  It can be expressed as $\langle resource_t,$ $property_t,$ $statement_t \rangle$. By leveraging the crisis-related information ontology, we are able to extract and capture the significant information from a new crisis situation and seamlessly incorporate it into an ontology-based knowledge base. This enables us to retrieve crisis-related information by querying a subgraph RDF, thereby making valuable contributions to the development of knowledge bases specifically tailored for crisis-related information. Our approach is inspired by the insightful research conducted by \cite{lengochal03675591,luyen2023}. Through this methodology, we have the capability to measure the similarity between specific triples or sets of triples, facilitating the analysis and comparison of crisis-related information.
	
	The measure of similarity between two specific triples is primarily based on the type of object present in each triple. In cases where the object type of the triple is qualitative, the semantic similarity of all three components of the triple, namely Qualitative Subjects, Predicates, and Objects (QSPO), will be measured as follows:
	
	\vspace{-1.0cm}
	\begin{equation}\vspace{-0.1cm}
		Sim_1(a_{s1},a_{s2}) = \frac{\sum_{i=1}^{k} \bar{S}(w_{1i}, a_{s2}) + \sum_{j=1}^{l} \bar{S}(w_{2j}, a_{s1})}{k + l}
	\end{equation}
	where $\bar{S}(w, a_{s})$ represents the similarity between a word $w$ and a QSPO $a_{s}$. With $\overrightarrow{w}$ is a embedding vector of the word $w$, the calculation of the function $\bar{S}(w, a_{s})$ is formally expressed as follows:

	\vspace{-0.4cm}
	\begin{equation}\vspace{-0.1cm}
		\bar{S}(w, a_{s}) = \max\limits_{w_q \in M} \bar{S}(\overrightarrow{w}, \overrightarrow{w_q})
		\vspace{-0.2cm}
	\end{equation}
	where $\overrightarrow{w_q} \in M=\{\overrightarrow{w_1}, \overrightarrow{w_2}, ..., \overrightarrow{w_z}\}$ refers to the word embedding vector of $a_s$, $z$ represents the number of words in $a_s$ and $q$ represents the index of the word in $M$ within the range of $0$ to $z$.
	If the objects in triples consist of numeric values, the comparison between these numbers is straightforward and involves measuring the distance between them. To compare two distinct objects, we employ the Euclidean distance as a metric because of its simplicity, interpretability, and wide applicability in various domains. Let's consider two objects denoted as $a_{o1}$ and $a_{o2}$, with respective vectors $\overrightarrow{a_{o1}} = \{o_{11}, o_{12}, ..., o_{1p}\}$ and $\overrightarrow{a_{o2}} =\{o_{21}, o_{22}, ..., o_{2p}\}$. In this context, their semantic similarity can be defined as follows:
	
	\vspace{-1.0cm}
	\begin{equation}
		\vspace{-0.2cm}
		Sim_2(\overrightarrow{a_{o1}},\overrightarrow{a_{o2}}) = \frac{1}{1 + \sqrt{\sum_{t=0}^{p}(o_{1t} - o_{2t})^2}}
	\end{equation}
Hence, the semantic similarity between two sets of triples, $C_1=\{a_{11}, a_{12}, ..., a_{1g}\}$ and $C_2=\{a_{21}, a_{22}, ..., a_{2g}\}$, is determined by comparing the similarity of each individual triple. The calculation is performed as follows:

\vspace{-1.0cm}
	\begin{equation}
		\vspace{-0.3cm}
		\begin{split}
			Sim(C_1, C_2) = \frac{1}{L}(\sum_{u=0}^{L} Sim_{1}(a_{1u},a_{2u})) \;+ \\  \frac{1}{H}(\sum_{v=0}^{H}Sim_{2}(\overrightarrow{a_{1v}},\overrightarrow{a_{2v}})) 
		\end{split}
	\end{equation}	 where $L$ denotes the number of triples containing qualitative objects, $H$ represents the number of triples containing quantitative objects, $g$ represents the number of triples in $C_1$ and $C_2$, while $u$ and $v$ denote the indices of the qualitative triple set and quantitative triple set, respectively.
	
	The measure of semantic similarity for different crisis situations entails the comparison of sets of triples representing crisis-related information in the ontology. This measure takes into account both quantitative and qualitative information present in the objects of the triples. For qualitative objects, the semantic similarity is calculated by utilizing numerical word vectors extracted from pre-trained word embeddings such as GloVe \citep{pennington2014glove}. This enables the computation of the semantic relatedness between words. Conversely, for quantitative objects, the similarity is evaluated by computing the Euclidean distance between their numerical vectors. The overall semantic similarity between two sets of triples is then determined by comparing the similarity of individual triples within the sets.
	\vspace{-0.2cm}
	\section{Experiments}\label{experiments}\vspace{-0.2cm}
	 In this section, we present the details of the data used for enriching the knowledge bases and provide an overview of the results obtained from applying our method for measuring semantic similarity on the enriched knowledge base. By showcasing the outcomes of our experiments, we aim to demonstrate the efficacy and practical applicability of our approach in identifying similar crisis situations.\vspace{-0.2cm}
	\subsection{Enrichment of the Knowledge Base}\vspace{-0.2cm}
	We present our initial experiments centered around extracting and enriching our ontology-based knowledge base in the domain of crisis-related information. To accomplish this, we harness the valuable EM-DAT database \cite{guha2016dat}. The EM-DAT database provides a comprehensive insights into various aspects of crises, including crucial details such as the dates and locations of events, the extent of their impact on affected populations (including figures for the number of individuals affected, injured, or deceased), economic losses incurred, and other relevant factors essential for comprehending the repercussions of these crises. Moreover, the database offers valuable information concerning the response and relief efforts implemented in the aftermath of these crisis situations. By incorporating data sourced from the EM-DAT database, we significantly enhance the depth and richness of our ontology-based knowledge base, thereby fostering a more comprehensive understanding of crisis-related information.
	
		\begin{figure}[th]
			\vspace{-0.3cm}
		\begin{tikzpicture}
			\begin{axis} [xbar ,
				ytick=data,
				tickwidth         = 0pt,
				y axis line style = { opacity = 0 },
				axis x line       = none,
				tick label style={font=\footnotesize},
				legend style={font=\footnotesize},
				label style={font=\footnotesize},
				width=.42\textwidth,
				height=7.3cm,style={align=right},
				bar width=12pt,
				xlabel={Number of crisis situations},
				yticklabels={Drought, Earthquake, Epidemic, Extreme-\\temperature, Flood, Industrial-\\accident, Landslide, Miscellaneous-\\accident,Storm, Transport- \\accident, Wildfire},
				nodes near coords,
				nodes near coords style={text=black,anchor=west,font=\footnotesize},visualization depends on=y \as \pgfplotspointy,
				every axis plot/.append style={fill}
				]
				\addplot  [dbscan,fill=dbscan] coordinates {
					(5,0) (4,1) (2,2) (22,3) (60,4) (14,5) (13,6) (24,7) (76,8) (54,9) (13,10)
				};
			\end{axis}
		\end{tikzpicture}
		\vspace{-0.3cm}
		\caption{Number of crisis situations that occurred in France, classified by crisis type, from 1903 to 2022}
		\label{fig_03}  
		\vspace{-0.6cm}
	\end{figure}
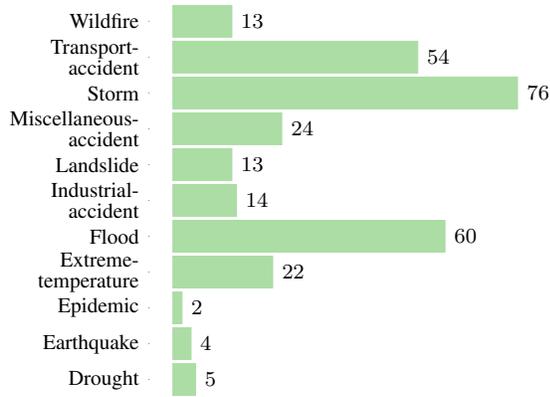
	
	The database contains a comprehensive record of 25,982 crisis situations that occurred worldwide from 1903 to 2022. These events encompass a diverse range of crisis types, including natural crisis such as earthquakes, floods, storms, droughts, and epidemics, as well as man-made crises such as industrial accidents, transportation incidents, and chemical spills. The coverage of the database spans across various countries, providing a global perspective on crisis occurrences. 
	In this work, we specifically focus on the crisis situations that have occurred in  France during this specified time period. Within this scope, we have identified approximately 287 distinct crisis situations that have taken place in France.  
		To provide a visual representation of the distribution, a statistical breakdown of these diverse types of crisis situations is visually presented in Figure \ref{fig_03}. This figure shows an informative overview of the different types of crises that have unfolded in France.

	By mining crisis situations from the EM DAT database and integrating them with a domain ontology for crisis-related information, our knowledge base pertaining to various crisis situations have experienced a substantial enhancement. This process of mining and integration has facilitated the expansion and refinement of the organization of crisis-related information by incorporating structured and standardized data from a wide array of sources. Consequently, we have achieved notable improvements in semantic information retrieval within the knowledge bases.
	
	The statistical data presented in table \ref{table_01} demonstrates the substantial progress made in the crisis-related information ontology. By extracting and enriching a total of 287 crisis situations that took place in France, we have substantially broadened the ontology's coverage, depth, and specificity. These advancements have greatly enhanced the ontology's value in facilitating similar crisis recognition. They enable the extraction, simulation, and learning from past experiences, offering valuable insights to stakeholders and decision-makers involved in crisis situations, both within France and globally. 
	
	
	\begin{table}[thb]
	\centering\vspace{-0.3cm}
	\caption{\label{table_01} Statistics on the Structure and Instances of the Crisis-Related Information Ontology }
	\label{tab: fonts}
	\begin{tabular}{|l|l|}
	\hline \bf Knowledge base statistics & \bf Quantitative\\\hline
 	Number of Class &  100 \\\hline
 	Number of Individual &  2540 \\\hline
 	Number of Object Property &  17 \\\hline
 	Number of Data Property &  40 \\\hline
 	Number of Statements &  100001 \\\hline
	\end{tabular}\vspace{-0.3cm}
	\end{table}

	{While RDF triples format is often associated with ontologies due to its ability to represent structured data and relationships, it's important to note that an RDF triples store can be created without incorporating an ontology. However, in our work, we have chosen to adopt an ontology-based approach to enhance the organization, semantics, and meaningful relationships within the domain of crisis-related information. This approach facilitates more robust analysis and reasoning in the subsequent stages of our work.} In the next section, we present the experiments we conducted to recognize similar crises using the achieved knowledge base.\vspace{-0.2cm}
	\subsection{Measuring Similarity between Crises}\vspace{-0.2cm}
	In the second experiment, we conducted SPARQL queries to retrieve the crisis-related information of each crisis from the ontology-based knowledge base in the form of RDF graphs. Utilizing our method for measuring semantic similarity between RDF graphs, we compared each crisis event with one another, resulting in a similarity score. A higher similarity score indicates a greater degree of similarity between two crisis situations.
	\begin{figure}[thb]
		\centering
		\centering\vspace{-0.4cm}
		\includegraphics[width=0.7\linewidth]{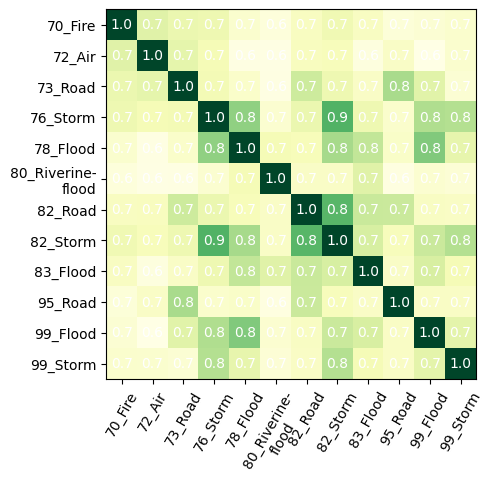}
		\vspace{-0.4cm}
		\caption[justification=justified]{Similarity scores of the twelve randomly extracted crisis situations. }
		\label{fig_04}       
		\centering\vspace{-0.4cm}
	\end{figure}
	Taking a closer look at the specific cases, we conducted a random comparison of twelve crisis situations by measuring similarity on their RDF triples. The main objective of this comparison was to evaluate the effectiveness of our similarity measure in capturing the similarities among these events. The computation results, as depicted in Figure \ref{fig_04}, clearly demonstrate that crises of a similar type tend to exhibit higher similarity scores. This observation suggests that the semantic characteristics encapsulated within the RDF triples greatly facilitate the identification and measure of similarities between crisis situations, particularly when they share comparable attributes or informative structures. For instance, when examining the specific events ``$80\_RiverineFlood$'' and ``$83\_Flood$'', we can observe that the former has a higher similarity score, as illustrated in Figure \ref{fig_04}. This further supports our claim that our approach effectively captures and quantifies the semantic relationships within the crisis-related information, thereby enhancing our ability to discern similarities between different crisis situations.
	\begin{figure}[thb]
		\vspace{-0.6cm}
		\centering
		\includegraphics[width=0.8\linewidth]{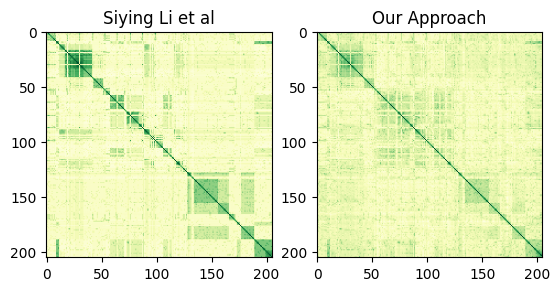}
		\caption[justification=justified]{Comparison heat chart illustrating the similarity scores of our approach and that of \cite{li2021ontology}}
		\label{fig_05}       
		\vspace{-0.5cm}
	\end{figure}
	
	To gain a comprehensive understanding of our similarity measure approach, we conducted a comparison with the approach proposed by \cite{li2021ontology}. Specifically, we compared 204 crisis situations that occurred in France using both approaches and visualized the results in the form of a heat chart, as depicted in Figure \ref{fig_05}. In general, our approach exhibits a smoother grouping of similar crises compared to Li's approach.
	In terms of numerical statistics, our approach yielded significantly higher scores, with 41,376 identified crisis similarities, whereas Li's approach only yielded 649. This indicates that our approach is more effective in capturing and quantifying the similarities between crisis situations, leading to a more comprehensive and accurate representation of crisis relationships. 	
	By conducting two comprehensive experiments focused on enriching the ontology-based knowledge base with crisis-related information and measuring semantic similarity between crises, our approach has demonstrated its effectiveness in identifying similar crises. The results obtained provide compelling evidence of the suitability of our approach in this purpose\vspace{-0.2cm}
	\section{Conclusion and Perspectives}\vspace{-0.2cm}
	In this paper, we conducted an investigation to recognize similar crises through the application of ontology-based knowledge mining. Our research encompasses two main aspects: (i) the development and enrichment of a crisis-related information ontology to facilitate the construction of the ontology-based knowledge base for crisis situations; and (ii) the application of ontology-based knowledge mining into the implementation of semantic similarity measures for computing relatedness between crisis situations. Through a systematic process cycle for identifying similar crisis situations, our experiments demonstrated the successful adaptation of existing crisis-related data into our ontology-based knowledge bases. Furthermore, our approach provided compelling evidence of its effectiveness in computing similarity scores between crisis situations. 
	It is essential to acknowledge the limitations of our work, particularly in practical applications where the knowledge bases should vary in terms of the information they contain regarding different crisis scenarios, actions, and corresponding communication. Therefore, it is crucial to proactively construct a comprehensive framework that leverages established standard ontologies for the knowledge base. 
	{ 
		Additionally, the similarity calculations rely on crisis instances from the ontology, which also shapes data organization. While our evaluation confirms semantic similarity's effectiveness, it is necessary to showcase how the ontology improves outcomes. Thus, we plan to compare results with and without ontology integration, highlighting its impact on similarity assessment.}
	\vspace{-0.2cm}
	\section*{Acknowledgement}\vspace{-0.2cm}
	
	This work was funded by the French Research Agency (ANR) and by the company Vivocaz under the project France Relance - preservation of R\&D employment (ANR-21-PRRD-0072-01).
	
	
	
	
	
	\bibliographystyle{apalike}
	\bibliography{references}

\begin{thebibliography}{}

\bibitem[Abdelghany et~al., 2019]{abdelghany2019agile}
Abdelghany, A.~S., Darwish, N.~R., and Hefni, H.~A. (2019).
\newblock An agile methodology for ontology development.
\newblock {\em International Journal of Intelligent Engineering and Systems},
  12(2):170--181.

\bibitem[Abiteboul et~al., 2011]{abiteboul_2011}
Abiteboul, S., Manolescu, I., Rigaux, P., Rousset, M.-C., and Senellart, P.
  (2011).
\newblock {\em Ontologies, RDF, and OWL}, page 143–170.
\newblock Cambridge University Press.

\bibitem[Arp et~al., 2015]{arp2015building}
Arp, R., Smith, B., and Spear, A.~D. (2015).
\newblock {\em Building ontologies with basic formal ontology}.
\newblock Mit Press.

\bibitem[Arru et~al., 2019]{arru:hal-02276437}
Arru, M., Negre, E., and Rosenthal-Sabroux, C. (2019).
\newblock {To Alert or Not to Alert? That Is the Question}.
\newblock In {\em {52nd Hawaii International Conference on System Sciences
  (HICSS 52)}}, Proceedings of the 52nd Hawaii International Conference on
  System Sciences | 2019, pages 649--658, Wailea, HI, United States.

\bibitem[Benaben et~al., 2020]{benaben2020ai}
Benaben, F., Fertier, A., Montarnal, A., Mu, W., Jiang, Z., Truptil, S.,
  Barthe-Delano{\"e}, A.-M., Lauras, M., Mace-Ramete, G., Wang, T., et~al.
  (2020).
\newblock An ai framework and a metamodel for collaborative situations:
  Application to crisis management contexts.
\newblock {\em Journal of Contingencies and Crisis Management}, 28(3):291--306.

\bibitem[Benaben et~al., 2008]{benaben2008metamodel}
Benaben, F., Hanachi, C., Lauras, M., Couget, P., and Chapurlat, V. (2008).
\newblock A metamodel and its ontology to guide crisis characterization and its
  collaborative management.
\newblock In {\em Proceedings of the 5th International Conference on
  Information Systems for Crisis Response and Management (ISCRAM), Washington,
  DC, USA, May}, pages 4--7.

\bibitem[Benaben et~al., 2016]{benaben2016metamodel}
Benaben, F., Lauras, M., Truptil, S., and Salatg{\'e}, N. (2016).
\newblock A metamodel for knowledge management in crisis management.
\newblock In {\em 2016 49th Hawaii International Conference on System Sciences
  (HICSS)}, pages 126--135. IEEE.

\bibitem[Boin et~al., 2020]{boin2020learning}
Boin, A., Lodge, M., and Luesink, M. (2020).
\newblock Learning from the covid-19 crisis: an initial analysis of national
  responses.
\newblock {\em Policy Design and Practice}, 3(3):189--204.

\bibitem[Boin and McConnell, 2007]{boin2007preparing}
Boin, A. and McConnell, A. (2007).
\newblock Preparing for critical infrastructure breakdowns: the limits of
  crisis management and the need for resilience.
\newblock {\em Journal of contingencies and crisis management}, 15(1):50--59.

\bibitem[Bowen and Lovari, 2020]{Bowen2020}
Bowen, S.~A. and Lovari, A. (2020).
\newblock {\em Crisis Management}, pages 1--10.
\newblock Springer International Publishing, Cham.

\bibitem[Bundy et~al., 2017]{bundy2017crises}
Bundy, J., Pfarrer, M.~D., Short, C.~E., and Coombs, W.~T. (2017).
\newblock Crises and crisis management: Integration, interpretation, and
  research development.
\newblock {\em Journal of management}, 43(6):1661--1692.

\bibitem[De~Nicola et~al., 2019]{de2019creative}
De~Nicola, A., Melchiori, M., and Villani, M.~L. (2019).
\newblock Creative design of emergency management scenarios driven by
  semantics: An application to smart cities.
\newblock {\em Information Systems}, 81:21--48.

\bibitem[Elmhadhbi et~al., 2020]{elmhadhbi2020promes}
Elmhadhbi, L., Karray, M.-H., Archim{\`e}de, B., Otte, J.~N., and Smith, B.
  (2020).
\newblock Promes: An ontology-based messaging service for semantically
  interoperable information exchange during disaster response.
\newblock {\em Journal of Contingencies and Crisis Management}, 28(3):324--338.

\bibitem[Gangemi et~al., 2002]{gangemi2002sweetening}
Gangemi, A., Guarino, N., Masolo, C., Oltramari, A., and Schneider, L. (2002).
\newblock Sweetening ontologies with dolce.
\newblock In {\em Knowledge Engineering and Knowledge Management: Ontologies
  and the Semantic Web: 13th International Conference, EKAW 2002 Sig{\"u}enza,
  Spain, October 1--4, 2002 Proceedings 13}, pages 166--181. Springer.

\bibitem[Gruber, 1993]{gruber1993translation}
Gruber, T.~R. (1993).
\newblock A translation approach to portable ontology specifications.
\newblock {\em Knowledge acquisition}, 5(2):199--220.

\bibitem[Guha-Sapir et~al., 2016]{guha2016dat}
Guha-Sapir, D., Below, R., and Hoyois, P. (2016).
\newblock Em-dat: the cred/ofda international disaster database.

\bibitem[Le et~al., 2022]{lengochal03675591}
Le, N.~L., Abel, M.-H., and Gouspillou, P. (2022).
\newblock {Apport des ontologies pour le calcul de la similarit{\'e}
  s{\'e}mantique au sein d'un syst{\`e}me de recommandation}.
\newblock In {\em {Ing{\'e}nierie des Connaissances (Ev{\`e}nement affili{\'e}
  {\`a} PFIA'22 Plate-Forme Intelligence Artificielle)}}, Saint-{\'E}tienne,
  France.

\bibitem[Le et~al., 2023a]{luyen2023}
Le, N.~L., Abel, M.-H., and Gouspillou, P. (2023a).
\newblock Improving semantic similarity measure within a recommender system
  based-on rdf graphs.
\newblock In {\em Proceedings of the 6th International Conference on
  Information Technology \& Systems}.

\bibitem[Le et~al., 2023b]{le:hal-03832226}
Le, N.~L., Zhong, J., Negre, E., and Abel, M.-H. (2023b).
\newblock {Constraint-based recommender system for crisis management
  simulations}.
\newblock In {\em {The 56th Hawaii International Conference on System
  Sciences}}, Hawaii, United States.

\bibitem[Li et~al., 2021]{li2021ontology}
Li, S., Abel, M.-H., and Negre, E. (2021).
\newblock Ontology-based semantic similarity in generating context-aware
  collaborator recommendations.
\newblock In {\em 2021 IEEE 24th International Conference on Computer Supported
  Cooperative Work in Design (CSCWD)}, pages 751--756. IEEE.

\bibitem[Meesters, 2021]{meesters2021crisis}
Meesters, K. (2021).
\newblock Crisis information management: From technological potential to
  societal impact.
\newblock {\em The New Common: How the COVID-19 Pandemic is Transforming
  Society}, pages 153--159.

\bibitem[Musen, 2015]{musen2015protege}
Musen, M.~A. (2015).
\newblock The prot{\'e}g{\'e} project: a look back and a look forward.
\newblock {\em AI matters}, 1(4):4--12.

\bibitem[Negre, 2013]{negre2013towards}
Negre, E. (2013).
\newblock Towards a knowledge (experience)-based recommender system for crisis
  management.
\newblock In {\em 2013 Eighth International Conference on P2P, Parallel, Grid,
  Cloud and Internet Computing}, pages 713--718. IEEE.

\bibitem[Niles and Pease, 2001]{niles2001towards}
Niles, I. and Pease, A. (2001).
\newblock Towards a standard upper ontology.
\newblock In {\em Proceedings of the international conference on Formal
  Ontology in Information Systems-Volume 2001}, pages 2--9.

\bibitem[Pennington et~al., 2014]{pennington2014glove}
Pennington, J., Socher, R., and Manning, C.~D. (2014).
\newblock Glove: Global vectors for word representation.
\newblock In {\em Proceedings of the 2014 conference on empirical methods in
  natural language processing (EMNLP)}.

\bibitem[Yu et~al., 2018]{yu2018risk}
Yu, F., Li, X.-Y., and Han, X.-S. (2018).
\newblock Risk response for urban water supply network using case-based
  reasoning during a natural disaster.
\newblock {\em Safety Science}.

\bibitem[‘t Hart, 2023]{t2023teaching}
‘t Hart, P. (2023).
\newblock Teaching crisis management before and after the pandemic: Personal
  reflections.
\newblock {\em Teaching Public Administration}, 41(1):72--81.

\end{thebibliography}
	
	
\end{document}